\documentclass[prl, twocolumn,
 amsmath, amssymb,
 aps,
floatfix,
]{revtex4-1}

\usepackage{graphicx}% Include figure files
\usepackage{dcolumn}% Align table columns on decimal point
\usepackage{bm}% bold math
\usepackage{amsmath,nccmath}

\usepackage[dvipsnames]{xcolor}
\usepackage{pifont}
\usepackage{hyperref}
\hypersetup{
    colorlinks=true,
    linkcolor=cyan,
    filecolor=magenta,      
    urlcolor=blue,
    citecolor=blue,
}

\newcommand{\perB}{%
  \resizebox{!}{1.2\fontcharht\font`0}{$\mkern-2mu\times\mkern-2mu$}% 
}

\begin{document}

\title{Stress-Induced Dinoflagellate Bioluminescence at the Single Cell Level}

\author{Maziyar Jalaal$^{1}$, Nico Schramma$^{1,2}$, Antoine Dode$^{1,3}$,
H{\'e}l{\`e}ne de Maleprade$^{1}$, \\ Christophe Raufaste$^{1,4}$, Raymond E. Goldstein$^{1}$
  \email{R.E.Goldstein@damtp.cam.ac.uk}}
\affiliation{$^{1}$Department of Applied Mathematics and Theoretical 
Physics, University of Cambridge, Cambridge CB3 0WA, 
United Kingdom\\
$^{2}$Max-Planck Institute for Dynamics and Self-Organization, G\"ottingen, Germany\\
$^{3}${\'E}cole Polytechnique, 91128 Palaiseau Cedex, France\\
$^{4}$Universit{\'e} C{\^o}te d'Azur, CNRS, Institut de Physique de Nice, CNRS, 06100 Nice, France
}%

\date{\today}% It is always \today, today,

\begin{abstract}
One of the characteristic features of many marine dinoflagellates is their bioluminescence, which lights up 
nighttime breaking waves or seawater sliced by a ship's prow.  While the internal biochemistry of light 
production by these microorganisms is well established, the manner by which fluid shear or mechanical forces 
trigger bioluminescence is still poorly understood.  We report controlled measurements of the relation between 
mechanical stress and light production at the single-cell level, using high-speed imaging of micropipette-held 
cells of the marine dinoflagellate \textit{Pyrocystis lunula} subjected to localized fluid flows or direct 
indentation.  We find a viscoelastic response in which light intensity depends on both the 
amplitude and rate of deformation, consistent with the action of stretch-activated ion channels. A 
phenomenological model captures the experimental observations.

\end{abstract}

\maketitle

Bioluminescence, the emission of light by living organisms, has been a source of commentary since ancient 
times \cite{Harvey}, from  Aristotle and Pliny the Elder, to Shakespeare and Darwin \cite{Darwin}, who, 
like countless mariners before him, observed of the sea, \textit{``... every part of the surface, which 
during the day is seen as foam, now glowed with a pale light. The vessel drove before her bows two billows 
of liquid phosphorus, and in her wake she was followed by a milky train. As far as the eye reached, the 
crest of every wave was bright,...''}. The glow Darwin observed arose most likely from bacteria or 
dinoflagellates, unicellular eukaryotes found worldwide in marine and freshwater environments.   

Bioluminescence is found in a large range of organisms, from fish to jellyfish, worms, fungi, and fireflies.  While 
discussion continues regarding the ecological significance of light production \cite{Haddock}, 
the \textit{internal} biochemical process that produces light is now well understood.  In the particular 
case of dinoflagellates \cite{Valiadi}, changes in intracellular calcium levels 
produce an action potential, opening voltage-gated proton channels in the 
membranes of organelles 
called \textit{scintillons}, lowering the pH within them \cite{fogel} and causing oxidation of the protein 
\textit{luciferin}, catalyzed by \textit{luciferase}.
Far less clear is the mechanism by which fluid motion triggers bioluminescence.

Early experiments on light emission  utilizing unquantified fluid stirring or bubbling \cite{biggley1969stimulable} 
were superseded over the past two decades by studies in the concentric cylinder geometry of Couette flow
\cite{latz1994excitation,Cussatlegras} and macroscopic contracting flows \cite{latz1995novel,latz2004hydrodynamic}. 
Subsequent experiments explored light production by cells carried by fluid flow against barriers 
in microfluidic chambers \cite{latz2008bioluminescent}, or subjected to the localized forces of an atomic 
force microscope  \cite{tesson2015mechanosensitivity}. From these have come  estimates of the 
\textcolor{black}{stress} needed to trigger light production.  Indeed, dinoflagellates can serve as probes of shear 
in fluid flows \cite{latz1994excitation,latz1995novel,Rohr1997use,foti2010use,hauslage2017pyrocystis,DeaneStokes}.
At the molecular level, 
biochemical interventions have suggested a role for 
%GTP-binding proteins \cite{Gproteins}, and for 
stretch-activated ion channels \cite{stretch-activated} \textemdash known to feature prominently in touch sensation \cite{touch} \textemdash leading to the hypothesis that 
fluid motion stretches cellular membranes, forcing channels open and starting the biochemical cascade that 
produces light. 
 
\begin{figure}[b]
\centering
\includegraphics[clip=true,width=0.90\columnwidth]{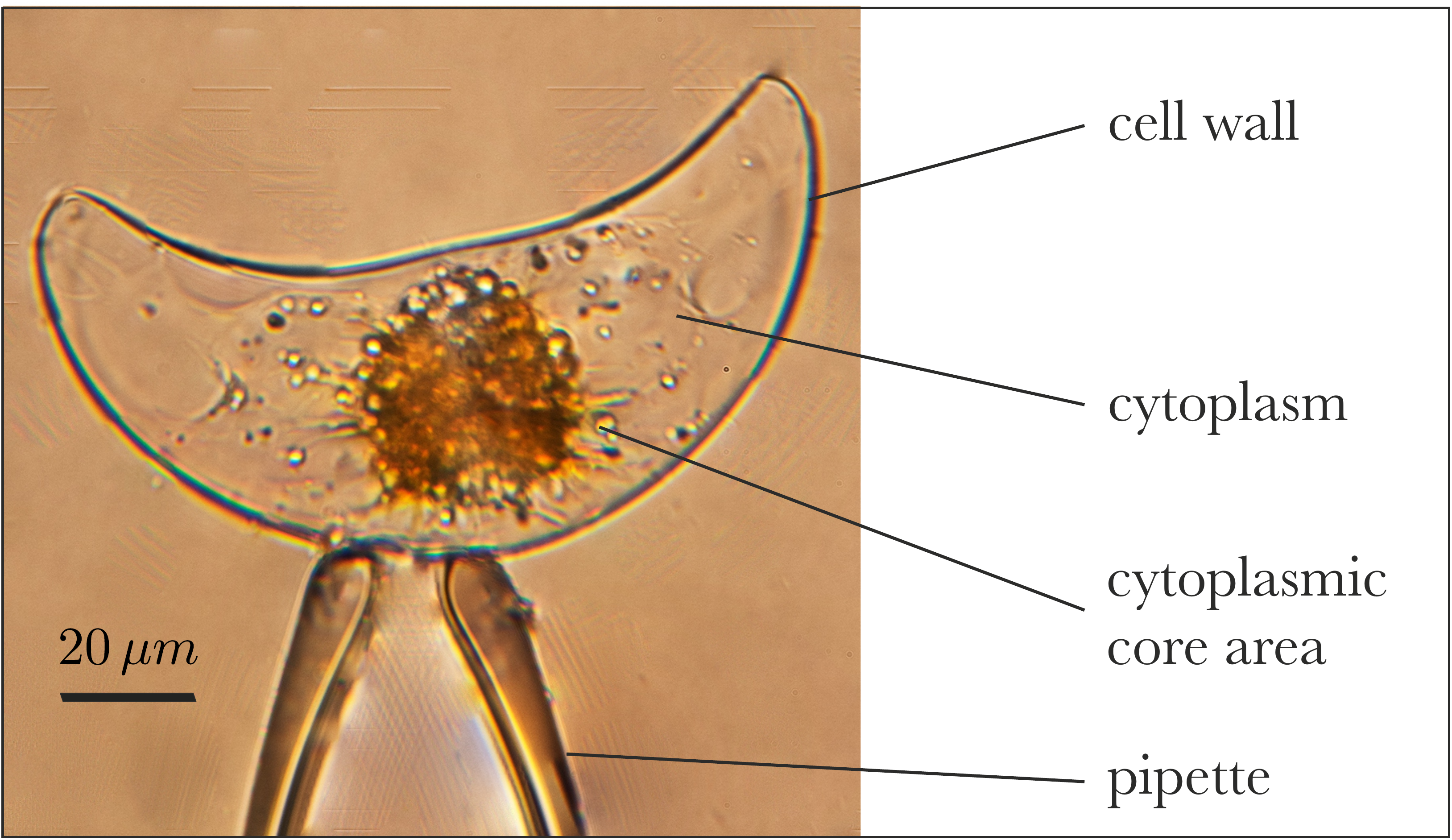}
\caption{The unicellular marine dinoflagellate \textit{Pyrocystis lunula}, held on a glass
micropipette. Chloroplasts (yellow/brown) are in the cytoplasmic core at night and the crescent-shaped cell 
wall encloses the cell.
\label{fig:lunul}}
\end{figure}

Here, as a first step toward an in-depth test of this mechanism, we study luminescence of single cells of the 
dinoflagellate \textit{Pyrocystis lunula} (Fig. \ref{fig:lunul}) induced by precise mechanical stimulation. 
The cellular response is found to be `viscoelastic', in that it depends not only on the 
amplitude of cell wall deformation but also on its rate.  A phenomenological model linking 
this behavior to light production provides
a quantitative account of these observations.

\begin{figure*}
\centering
\includegraphics[width=1.95\columnwidth]{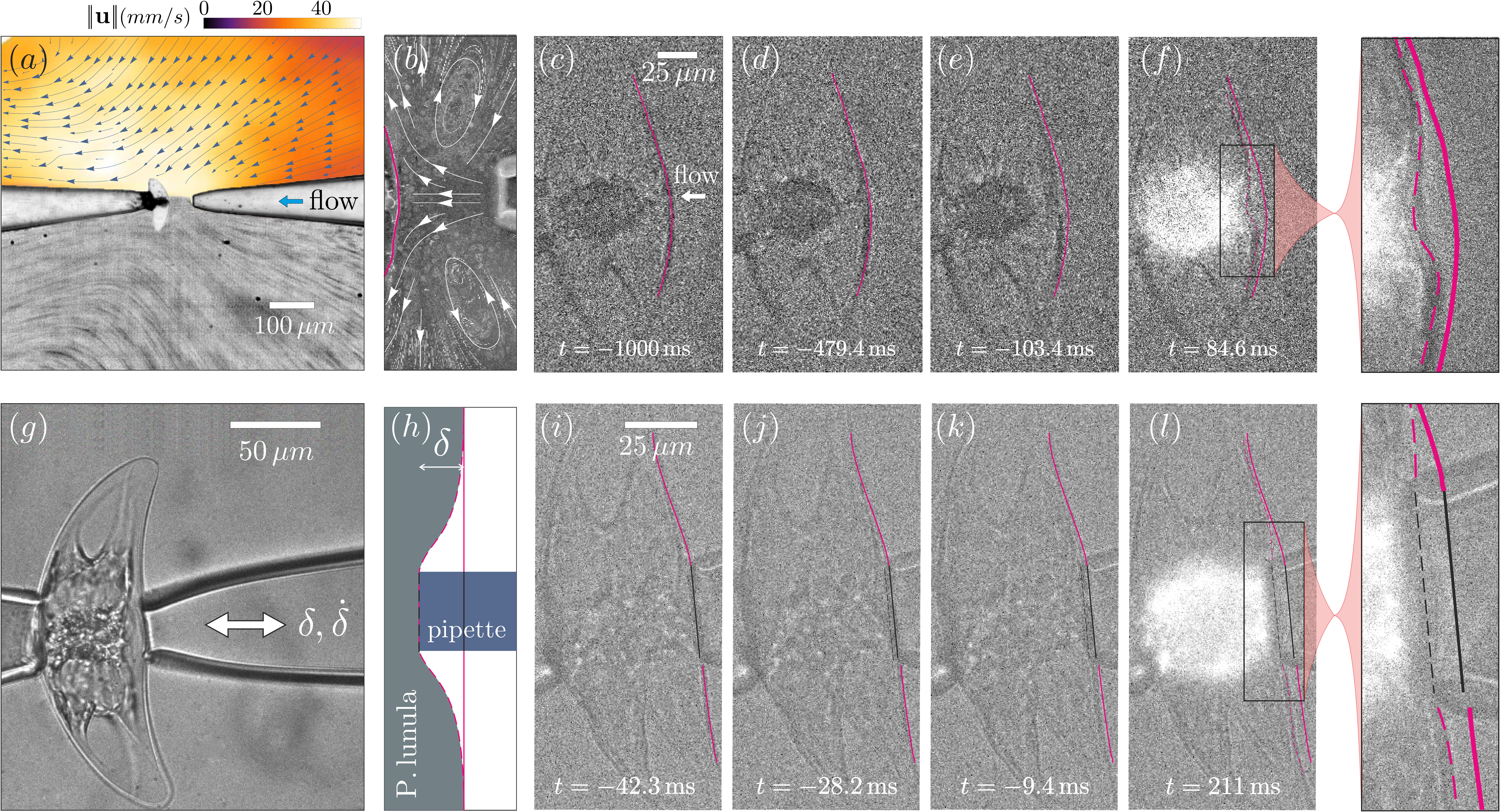}
\caption{Light production by \textit{P. Lunula} under fluid and mechanical stimulation. (a) Stimulation 
by fluid flow; color map in upper half indicates flow speed, lower half is a streak image of 
tracer particles. 
(b) Particle tracking of flow lines near cell surface. (c-f) Cell deformation due to fluid flow 
and the consequent 
light production. (g,h) Second protocol, in which a
cell is deformed under direct contact by a second pipette. (i-l) Light production 
triggered by mechanical deformation.  All times indicated are with respect to the start of light emission.}
\label{fig:Exp}
\end{figure*}
 
\textit{P. lunula} is an excellent organism for the study of bioluminescence because
its large size ($\sim\! 40\,\mu$m in diameter and 
$\sim\! 130\,\mu$m in length), lack of 
motility as an 
adult, rigid external cell wall and negative buoyancy all facilitate micromanipulation.  Its relative
transparency and featureless surface allow for high-resolution imaging. As model organisms, 
dinoflagellates have been studied from a variety of complementary perspectives \cite{lunula_model}.

Cultures of \textit{P. lunula} (Sch\"utt) obtained from CCAP \cite{CCAP} were grown in L1 medium \cite{guillard1993stichochrysis} at 
$20^\circ$C in an incubator on a 12h/12h light/dark cycle. The 
bioluminescence of \textit{P. lunula} is under circadian 
regulation \cite{swift1967bioluminescence,colepicolo1993circadian} and occurs only during the night. 
All experiments were performed between hours $3-5$ of the nocturnal phase.
An sCMOS camera 
(Prime 95B, Photometrics) imaged cells through a Nikon 63\perB \, water-immersion
objective on a Nikon TE2000 inverted microscope. Cells were kept in a $500\,\mu$L 
chamber that allows access by two antiparallel micropipettes held on
multi-axis micromanipulators 
(Patchstar, Scientifica, UK) (see Supplemental Material \cite{suppmat}), and kept undisturbed for several hours prior to 
stimulation.  
Upon aspiration on the first pipette, cells typically flash 
once \cite{firstflash}. Care was taken to achieve consistent positioning
of cells for uniformity of light measurements 
(Video 1 \cite{suppmat}). 

The second pipette applies mechanical stimulation in either of 
two protocols. The first directs a submerged jet of growth medium at the cell,
controlled by a syringe pump (PHD2000, Harvard Apparatus) and characterized using Particle 
Image Velocimetry (PIV) and particle tracking, as in Figs. \ref{fig:Exp}a-f.
Typical flow rates through the micropipette were on the order of $1$ ml/h, exiting a tip 
of radius $\sim 10\,\mu$m, yielding maximum jet speeds $U$ up to $1$\,m/s.  
With $\nu=\eta/\rho =1$\,mm$^2$/s 
the kinematic viscosity of water and $\ell\sim 0.02$\,mm the lateral size of the organism, the
Reynolds number is $Re=U\ell/\nu\sim 20$, consistent with prior 
studies in macroscopic flows 
\cite{latz1994excitation,latz1995novel,latz2004hydrodynamic}, which utilized the
apparatus scale (mm) for reference.
In the second protocol, a cell is held between the two pipettes, and mechanical deformation 
is imposed by displacement of the second. Using the 
micromanipulators and a computer-controlled translation
stage (DDS220/M, Thorlabs), the deformation $\delta$ and deformation rate $\dot{\delta}$ could 
be independently varied (Figs. \ref{fig:Exp}g-l). 

\begin{figure*}
\centering
\includegraphics[width=1.0\textwidth]{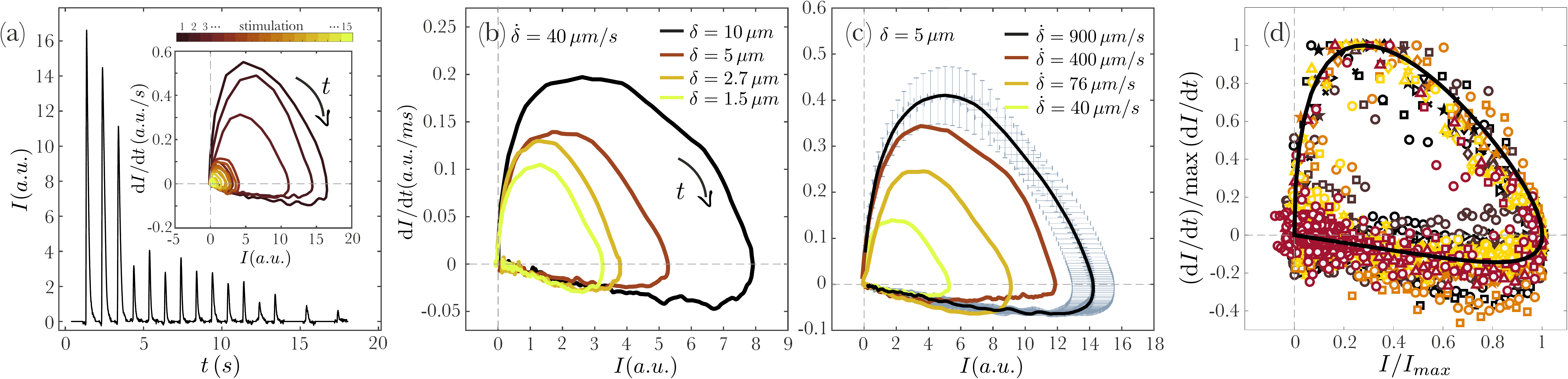}
\caption{Dynamics of light production following mechanical stimulation. (a) Response of a 
cell to repeated deformation with $\delta_f = 10\,\mu$m and $\dot{\delta} = 76\,\mu$m/s. Inset: loops
in $I-dI/dt$ plane for successive flashes. (b) Loops at fixed $\dot\delta$ and varying $\delta_f$ for first flashes. 
(c) As in (b), but for fixed
$\delta_f$ and varying $\dot\delta$.  Standard errors are shown for outermost data.
(d) Master plot of data, normalized by maximum intensities and rates.  Circles (squares) are data in b (c). Black
curve is result of model in \eqref{model2} and \eqref{model1}.}
\label{fig:loops}
\end{figure*}

A key observation within the first protocol is that cells do not flash unless the 
imposed fluid  pressure is high enough to deform the cell membrane sufficiently (Fig. \ref{fig:Exp}f).
For these submerged jet flows, the fluid stress $\Sigma_f\sim \rho U^2$ can be estimated to reach
$\sim\! 10^3$\, Pa, which is the same order as in prior macroscopic 
experiments \cite{latz1994excitation,latz1995novel,latz2004hydrodynamic}.  It can be seen 
from Fig. \ref{fig:Exp}f that the lateral scale 
$\xi$ of cell wall deformations is $\sim 30\,\mu$m, and we estimate the fluid force exerted
at the site of deformation as $F_f \sim \Sigma_f\xi^2\sim 0.1\,\mu$N.  
More quantitatively, using PIV of the flow field and
finite-element calculations of the flow from a pipette \cite{suppmat} we find from study of 
$35$ cells that
the threshold for light production is broadly distributed, with a peak at $0.10\pm 0.02\,\mu$N.  

It is not clear \textit{a priori} whether the deformations in Figs. \ref{fig:Exp}c-f 
are resisted by the cell wall alone or also by the cytoplasm. The wall has 
a tough outer layer above a region of cellulose fibrils \cite{Fensome,SwiftRemsen,SeoFritz},
with a thickness $d\sim 200-400$\, nm: AFM studies \cite{tesson2015mechanosensitivity}
show a Young's modulus $E \sim 1$\,MPa.
During asexual reproduction, the cellular contents 
pull away from the wall and eventually exit it through a hole, leaving behind
a rigid shell with the characteristic crescent moon shape \cite{SwiftDurbin}. Thus, 
the wall is not only imprinted with that shape, but is much more rigid than the
plasma membrane and significantly more rigid than the cytoplasm \cite{tesson2015mechanosensitivity}. 

Deformations of such curved structures induced by localized forces involve
bending and stretching of the wall.  With $\ell$ the radius of curvature of the undeformed cell
wall, a standard analysis \cite{LandauLifshitz} gives the indentation force
$F\sim Ed^2\delta/\ell$. Balancing this against the fluid force $\rho U^2\xi^2$
we find the strain $\varepsilon\equiv \delta/\ell \sim \left(\rho U^2/E\right)\left(\xi/d\right)^2$. 
From the estimates above, we have $\rho U^2/E\sim 10^{-4}$, and $\xi/d\sim 50-100$, 
so $\varepsilon$ is of the magnitude observed.  

In the natural setting of marine bioluminescence and in laboratory studies 
of dilute suspensions, light production can arise purely 
from flow itself, without contact between dinoflagellates.  Nevertheless,
there are conceptual and methodological advantages to studying bioluminescence by direct 
mechanical contact, especially due to the natural compliance of cells aspirated by a single micropipette.  
Chief among these is the ability to control the deformation and
deformation rate, which are the most natural variables for quantification of membrane stretching
and bending.  As seen in Fig. \ref{fig:Exp}i-l, imposing deformations similar to those 
achieved with the fluid flow also produces bioluminescence, highlighting the role of cell 
membrane deformation in mechanosensing. 

In our protocol for deformations, $\delta$ is increased at a constant rate 
$\dot\delta$ for a time $t_f$ to a final value $\delta_f$ (\textit{loading}), after which it was held fixed until 
any light production ceases, then returned to zero (\textit{unloading}).  We observe generally that if 
light is produced during loading, it is also produced during unloading. 
Experiments were performed for $\delta_f \in [1, 10]\, \mu$m and 
$\dot{\delta} \in [10, 900]\, \mu$m/s, with eight to twelve replicates (cells) for each 
data point. We repeated the given deformation protocol on the same cell 
(with sufficient rest intervals in between) until the cell ceases bioluminescence.  Reported
values of light intensity $I(t)$ are those integrated over the entire cell.

Figure \ref{fig:loops}a shows the light flashes from $15$ stimulations 
of a single cell. 
With each deformation, $I(t)$ first rises rapidly and then decays on a longer 
time scale. Apart from a decreasing overall magnitude with successive flashes, 
the shape of the signal remains nearly constant.  The eventual 
loss of bioluminescence most likely arises from exhaustion of the luciferin pool \cite{luciferin_loss}. 
%recovery takes several hours.
The inset shows the corresponding phase portraits of the flashes in the $I-dI/dt$ 
plane, where the similarity of successive signals can clearly be seen.

\begin{figure}[t]
\centering
\includegraphics[clip=true,width=0.95\columnwidth]{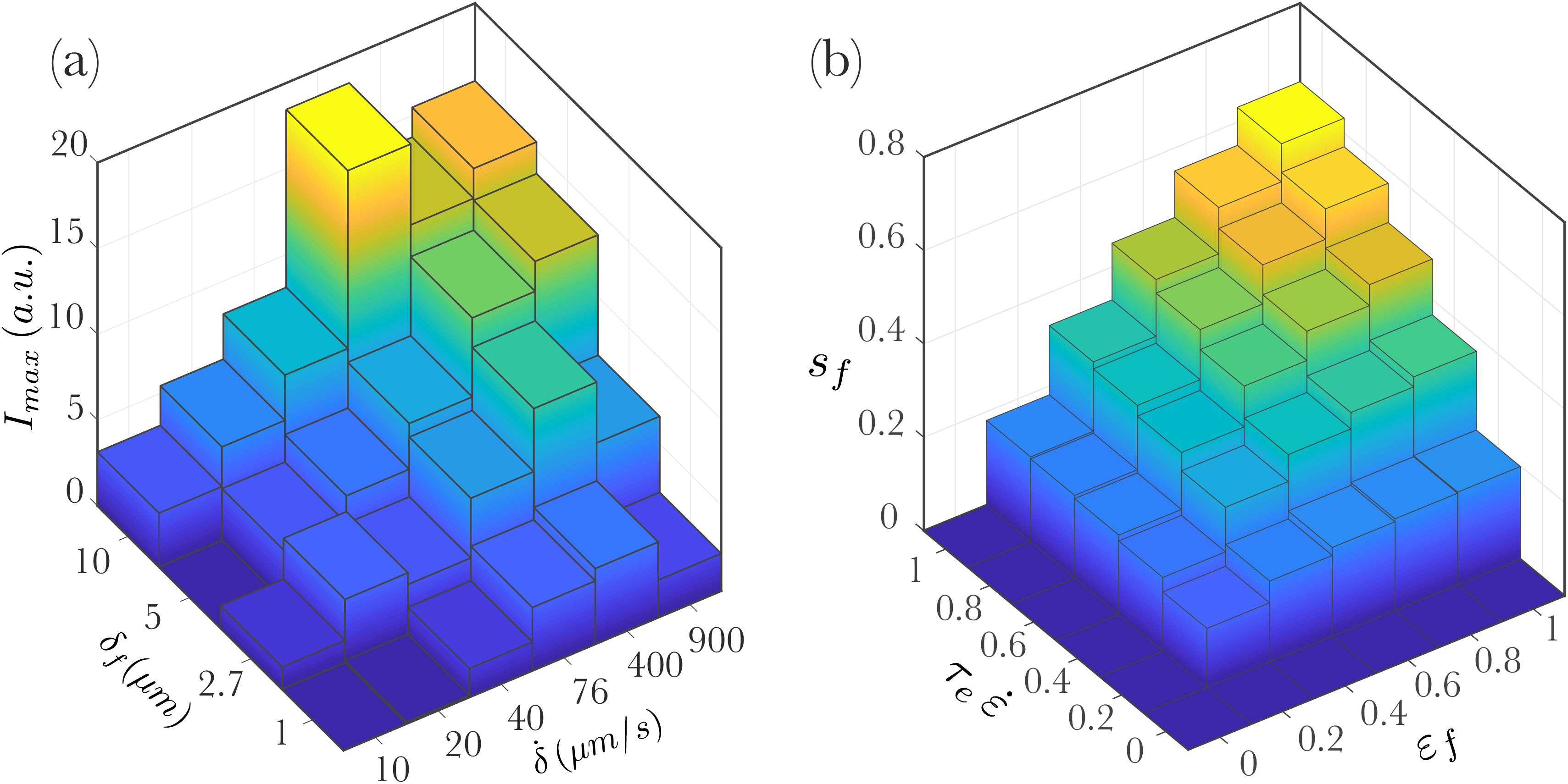}
\caption{Dependence of light production on deformation and rate. (a) Histogram of maximum  
intensity. 
Note nonuniform grid. (b) Variation of signal strength $s_f$ predicted
by phenomenological model, as a function of deformation and rate.}
\label{fig:Imax}
\end{figure}

Focusing on the first flashes, experiments with different $\delta_f$ and $\dot\delta$ 
reveal the systematics of
light production. Figures \ref{fig:loops}b$\&$c 
show that for a given rate, larger deformations produce more light, as do 
higher rates at a given deformation. 
Interestingly, the shape of the signals remains the same not only between different cells but 
also for different mechanical stimulations; normalizing the phase portraits with respect 
to their maxima yields a universal shape of the signal (Fig. \ref{fig:loops}d). 
We summarize the results of all experiments in Figure \ref{fig:Imax}a, showing the variation of maximum 
light intensity (averaged over all the first flashes) as a function of $\delta_f$ and $\dot{\delta}$;
light production is maximized when the cell is highly deformed at high speed. 

The influence of deformation and rate are
suggestive of viscoelastic properties. At a phenomenological level, we thus consider 
a Maxwell-like model that relates the signal $s(t)$ that triggers light production to the
strain $\varepsilon$,
\begin{align} 
\dot{s} + \tau_e^{-1}s = \dot{\varepsilon}~,
\label{model2}
\end{align}
where $\tau_e$ is a relaxation time.
For a given $\delta$, 
if the deformation time scale is much smaller than $\tau_e$, the membrane does not have time 
to re-arrange (the large Deborah number regime in rheology), while for slow deformations 
the membrane has time to relax.  As seen in Figs. \ref{fig:Exp}i-l and Videos 2 \& 3 
\cite{suppmat}, bioluminescence occurs \textit{during} 
loading, a feature that suggests
$\tau_e$ is comparable to the flash rise time.
Integrating \eqref{model2} up to $t_f$, we obtain the signal $s_f$ at the end of loading in terms of the 
final strain $\varepsilon_f\equiv\delta_f/\ell$ and scaled strain rate $\dot\varepsilon\tau_e$, 
\begin{equation}
    s_f=\dot\varepsilon \tau_e \left(1-{\rm e}^{-\varepsilon_f/\dot\varepsilon\tau_e}\right)~.
    \label{sf}
\end{equation}
As seen in Fig. \ref{fig:Imax}b, the peak response occurs when both the final
strain and strain rate are large, as observed experimentally. 
The linear 
relationship between $s$ and $\varepsilon$ embodied in \eqref{model2} can not continue to 
be valid at large strains or strain rates; eventually, the signal must saturate when all available channels
open to their maximum.  This is consistent with the data in Fig. \ref{fig:Imax}a at the 
highest rates, where experimentally $\varepsilon \sim 0.25$.

Although light production is triggered internally by an action potential\textemdash which arises from 
nonlinear, \textit{excitable} dynamics\textemdash analysis of the flashes \cite{suppmat} indicates 
a time course much like that of 
two coupled capacitors charging and discharging on different time scales. 
Such linear dynamics have figured in a variety of 
contexts, including calcium oscillations \cite{Goldbeter}, bacterial chemotaxis \cite{Othmer}, and 
algal phototaxis \cite{Fidelity}, and take the form of coupled equations for the observable
(here, the light intensity $I$) which reacts to the signal $s$ on a short time $\tau_r$ 
and the hidden biochemical process $h$ which resets the system on a longer time $\tau_a$.
For light triggered by stretch-activated ion channels, the signal $s$ might be 
the influx of calcium resulting from the opening of channels. 
Adopting units in which $I,h,s$ are dimensionless, the simplest model is
\begin{subequations}
\begin{align} 
\tau_r \, \dot{I} &= s - h - I\,,\\
\tau_a \, \dot{h} &= s - h\,.
\end{align}
\label{model1}
\end{subequations}
\vskip -0.5cm

\indent  Starting from the fixed point $(I=0,h=0)$ for $s=0$, if the signal is turned on abruptly
then $I$ will respond on a time scale $\tau_r$,
exponentially approaching $s-h\simeq s$.  Then, as $h$ evolves toward $s$ over the longer adaptation time scale $\tau_a$, 
$I$ will relax toward $s-h\simeq 0$, completing a flash.  
It follows from \eqref{model1} that a discontinuous initial $s$ creates a discontinuous
$\dot I$, whereas the loops in Fig. \ref{fig:loops} show
smooth behaviour in that early regime ($I, \dot{I}\gtrsim0$); this smoothing arises 
directly from the Maxwell model \eqref{model2} for the signal.
The parsimony of the linear model \eqref{model1} comes at a cost, for it fails at very high ramp rates
when $\dot\varepsilon$ switches to zero within the flash period and both $s$ and $I$ would 
adjust accordingly, contrary to observations. In a more complex, excitable model, the flash, once 
triggered, would 
thereafter be insensitive to the signal.

As the entire system \eqref{model2} and \eqref{model1} is linear, it can be solved exactly 
\cite{suppmat}, thus enabling a global fit to the parameters.  
We compare the theoretical results with the 
normalized experimental data in Fig. \ref{fig:loops}d, where we see good agreement with the
common loop structure. From the fits across all data, we find 
common time scales $\tau_e \approx 0.027\,$s, $\tau_r \approx 0.012\,$s,
and $\tau_a \approx 0.14\,$s, the last of which is comparable to the pulse decay time found in
earlier experiments with mechanical stimulation \cite{firstflash}, and can be read off
directly from the late-time dynamics of the loops in Figs. \ref{fig:loops}b\&c, where
$\dot I \sim -I/\tau_a$ \cite{suppmat}. 
These values suggest comparable time scales 
of membrane/channel viscoelasticity and biochemical actuation, both much shorter than the decay of 
light flashes.

\begin{figure}[b]
\centering
\includegraphics[clip=true,width=1.0\columnwidth]{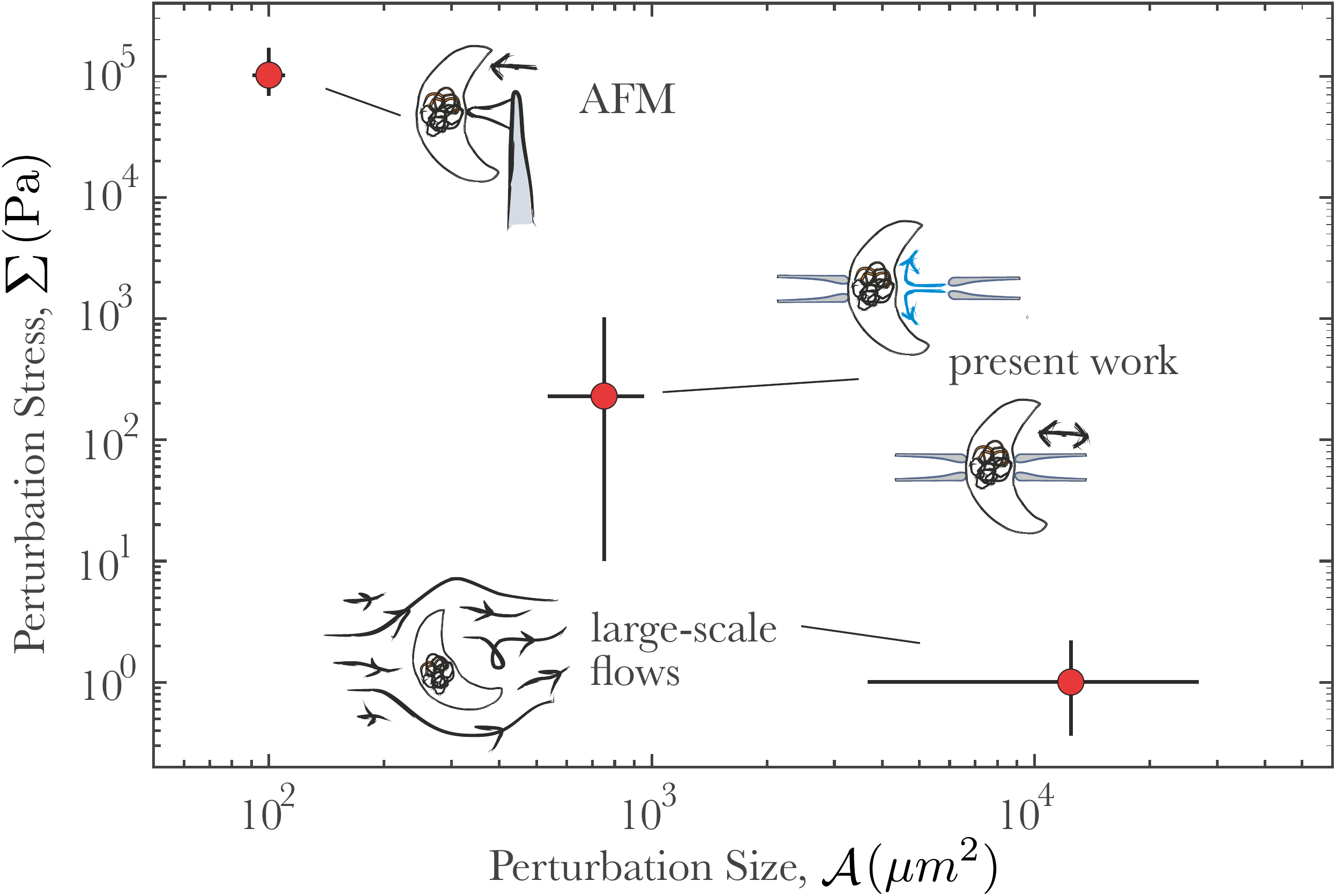}
\caption{Perturbation stress versus perturbation area for three kinds of experiments on 
dinoflagellates.  Atomic force measurements on \textit{P. lunula} 
are from 
\cite{tesson2015mechanosensitivity}, while macroscopic measurements include
Taylor-Couette \cite{latz1994excitation,Cussatlegras} and
contracting flows \cite{latz1995novel,latz2004hydrodynamic} on \textit{P. lunula} 
and similar dinoflagellates.}
\label{fig:PertStress}
\end{figure}

With the results described here, the generation of bioluminescence has now been explored with
techniques ranging from atomic force microscope cantilevers with attached microspheres 
indenting cells in highly localized areas, to fluid jets and micropipette indentation on intermediate
length scales, and finally to macroscopic flows that produce shear stresses across the entire
cell body. Figure \ref{fig:PertStress} considers all of these experiments together, organized by
the perturbative stress $\Sigma$ found necessary to produce light and the area $\mathcal{A}\equiv \xi^2$ 
over which that stress was applied.  We see a clear trend; the smaller the perturbation area, the larger the 
force required. This result suggests that the production of a given amount of 
light, through the triggering effects of stretch-activated ion channels on intracellular action potentials, 
can be achieved through
the action of many channels weakly activated or a small number strongly activated.
With an eye toward connecting the present results to the familiar marine context of light
production, it is thus of interest to understand more quantitatively the distribution 
of forces over the entire cell body in strong shear flows \cite{theory} and how those forces activate
ion channels to produce light.  Likewise, the possible ecological significance of the great
range of possible excitation scales illustrated in Fig. \ref{fig:PertStress} remains to
be explored.

\begin{acknowledgments}
We are grateful to Michael I. Latz for invaluable assistance at an early stage of this work, 
particularly with regard to culturing dinoflagellates, and thank Adrian Barbrook, Martin Chalfie, Michael Gomez, 
Tulle Hazelrigg, Chris Howe,
Caroline Kemp, Eric Lauga, Benjamin Mauroy, Carola Seyfert,
and Albane Th{\'e}ry for important discussions.
This work was supported in part by the Gordon and Betty Moore Foundation (Grant 7523) and the 
Schlumberger Chair Fund. CR acknowledges support by the French government, through the UCA$^{\mbox{JEDI}}$ 
Investments in the Future project of the National Research Agency (ANR) (ANR-15-IDEX-01).
\end{acknowledgments}

% \end{document}

% \documentclass[onecolumn,superscriptaddress,floatfix,preprintnumbers]{revtex4}
% \usepackage{graphics,amssymb,amsmath,epsfig,color}
% \usepackage{graphicx}
% \usepackage{comment}

% \begin{document}

\vfil
\eject

\widetext

\section{Supplemental Material}

\setcounter{equation}{0}
\setcounter{figure}{0}
\setcounter{table}{0}
\setcounter{page}{1}
\makeatletter
\renewcommand{\theequation}{S\arabic{equation}}
\renewcommand{\thefigure}{S\arabic{figure}}
\renewcommand{\bibnumfmt}[1]{[S#1]}
\renewcommand{\citenumfont}[1]{S#1}
%%%%%%%%%% Prefix a "S" to all equations, figures, tables and reset the counter %%%%%%%%%%

\section{Experimental Setup }

\begin{figure*}[b]
\centering
\includegraphics[clip=true,width=0.75\columnwidth]{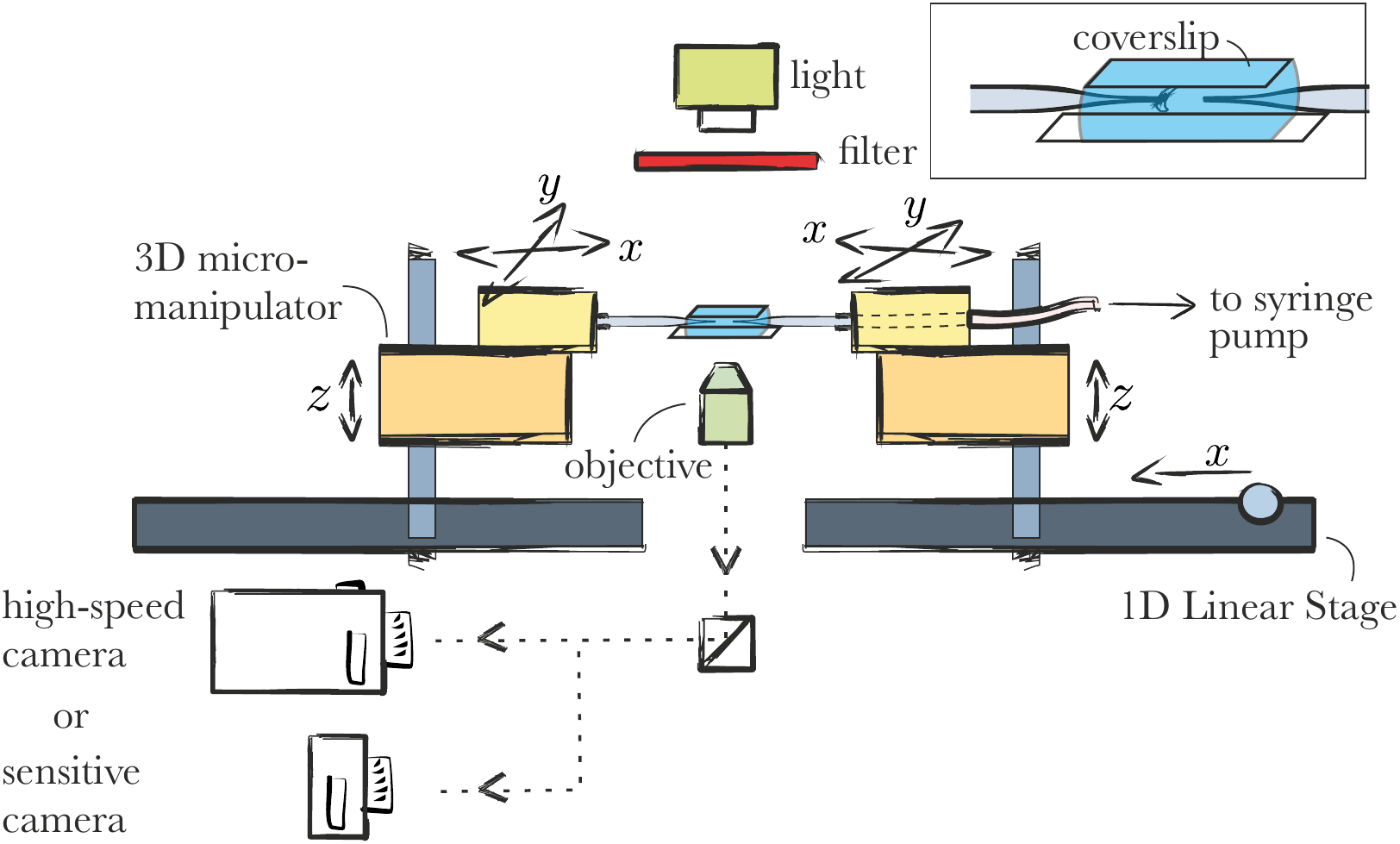}
\caption{Schematic of experimental setup to study bioluminescence 
produced by single dinoflagellates under controlled stresses. 
\label{fig:setup}}
\end{figure*}

The experimental setup, shown schematically in Figure \ref{fig:setup}, 
consists of a microscope for visualization and positioning 
systems to control the two pipettes. 
All experiments were conducted in 
a darkened room.
The white light illumination of the microscope (Nikon TE2000 ) 
was kept to a minimum and sent through a red long-pass filter 
($620$ nm, Knight Optical, UK) to avoid disruption of the 
night phase of the dinoflagellates and to allow a greater dynamic range 
in capturing the bioluminescence. That
background intensity was controlled in all experiments for uniformity. 

Pipettes were positioned with 3D micromanipulators (Patchstar, Scientifica).
For small deformation rate experiments, we used a Thorlabs 
1D Direct Drive Linear Stage (DDS220/M) to control the motion. All stages 
were programmed with their native software. The pipettes 
were connected to syringes with stiff tubing and fluid flow through 
them could also be used to position the cells (see below). 
In the flow experiments, we used a syringe pump (PHD2000, 
Harvard Apparatus) at a constant rate.
 The test section was a chamber whose top and bottom were coverslips, 
held apart by $\sim 2$ mm plastic spacers (see inset of Fig.
\ref{fig:setup}).  As \textit{P. lunula} has a very characteristic 
three dimensional geometry, for consistency, we held the cells the 
same orientation within the chamber in all experiments. 

\begin{figure}[t]
\centering
\includegraphics[clip=true,width=0.9\columnwidth]{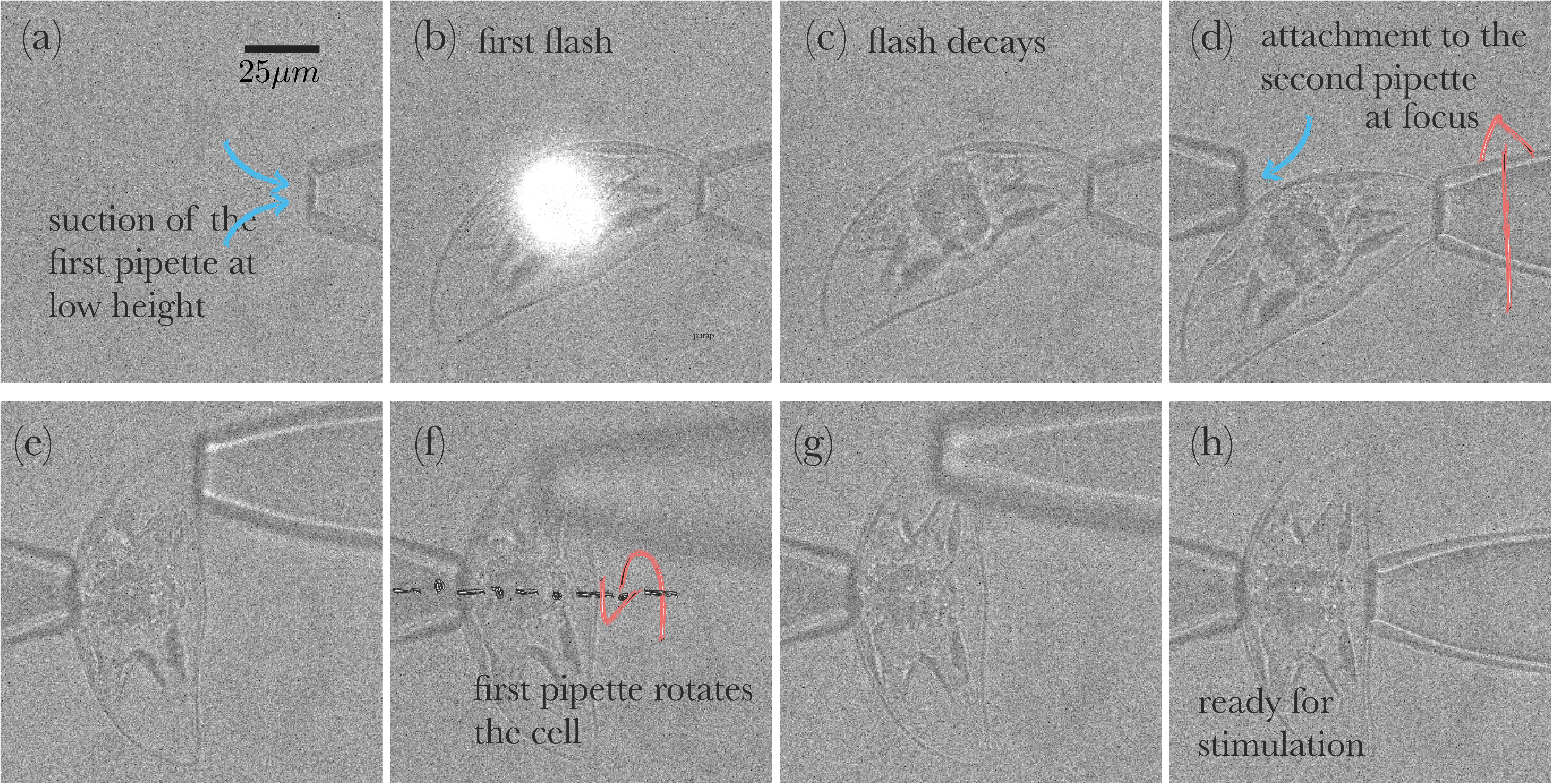}
\caption{Manual positioning of a cell prior to main measurements. (a) The 
cell is initially drawn up from the bottom cover glass using gentle flow 
suction.  It nearly always aspirated from one of its pointy ends. 
The cell flashes once in this process (b) and the light decays (c). 
The pipettes are raised within the sample chamber to be far from 
the top and bottom chamber surfaces. (d) Using the joystick controllers 
of the micromanipulators, the cell is placed on the other pipette and then 
held using gentle suction (e).  (f,g) The cell is rotated so that 
the largest area is exposed to the camera. (h) Finally, by placing the 
cell between the pipettes, indentation experiments can be performed.
\label{fig:position}}
\end{figure}

As cells of \textit{P. lunula} are negatively buoyant, they settle to 
the bottom coverslip of the sample chamber.  
Cells were positioned manually with the use of joystick controllers and 
gentle suction of the flow, as illustrated in Figure \ref{fig:position}.
The main bioluminescence experiments were recorded with a Prime 95B 
sCMOS camera (Photometrics). The high sensitivity of the camera allowed 
for measurements at low light condition but relatively high recording 
speed.  For the PIV and particle tracking experiments, we used a 
high-speed camera (Phantom v311). Figure 1 of the main text was 
captured using a Nikon D810 DSLR with Differential Interference Contrast (DIC) microscopy.

\section{Solution of the model}

Equations 1, 3a and 3b are linear ODEs which can be solved 
exactly. As described in the main text, we take here the simplest case in which
the light flash occurs within the ramp period, and therefore confine the discussion
to times $t<t_f$, during which the rate of strain $\dot{\varepsilon}$ is 
constant.  From the three time constants ($\tau_r,\tau_e,\tau_a$) we find 
$\tau_a$ to be by far the largest, and thus define the two ratios $\lambda,\rho < 1$,
\begin{equation}
     \lambda=\frac{\tau_e}{\tau_a}\,, \ \ \ \ \rho=\frac{\tau_r}{\tau_a}\,.
\end{equation}
Then, from (1) the signal is
\begin{equation}
s(t) =  \dot{\varepsilon}\tau_e \left( 1 - e^{-t/\tau_e} \right)\,.
\label{eq:s}
\end{equation}
As $t \rightarrow 0$, $s \sim \dot{\varepsilon} t + \cdots$, while at long times
$s$ approaches $\dot{\varepsilon}\tau_e$.  If we set $t=t_f$ and note that 
$t_f \equiv  \varepsilon_f / \dot{\varepsilon}$, we obtain (2) in the main text.
Substituting \eqref{eq:s} into (3b) and solving for $h$, we find
\begin{equation}
h = \frac{\dot{\varepsilon}\tau_e}{1-\lambda}\left[1 - e^{-t/\tau_a} 
-\lambda \left(1-e^{-t/\tau_e}\right)\right]\,.
\label{eq:h}
\end{equation}
which varies as $\dot{\varepsilon} t^2/2\, \tau_a$ as $t \rightarrow 0$ and, as with 
$s$, approaches $\dot{\varepsilon}\tau_e$ for long times.

\begin{figure}[t]
\centering
\includegraphics[clip=true,width=1\columnwidth]{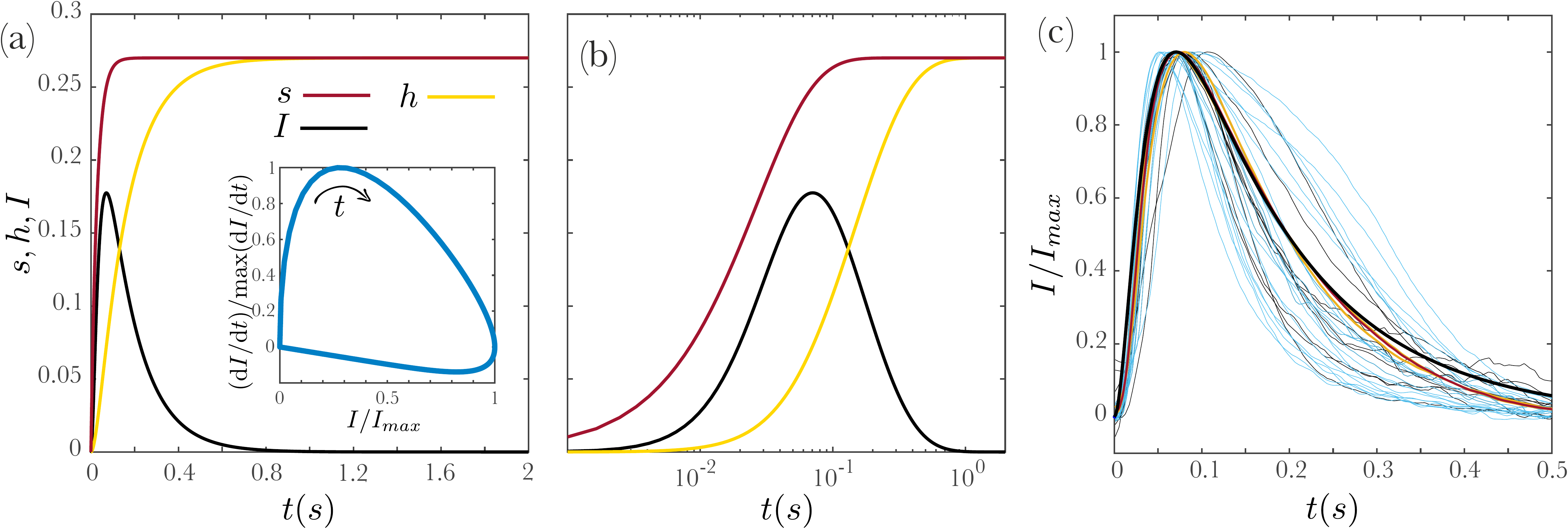}
\caption{a) Plots of $s,h,$ and $I$ for $\tau_r=0.012$, $\tau_a=0.1$4, $\tau_e=0.027$, and $\dot{\delta} = 500 \mu$m/s. The inset shows the phase portrait of the intensity signal. b) The same as (a) in a linear-log scale to highlight the early time behaviour. c) Intensity signals corresponding to the results shown in figure 3. Black and blue thin lines show the raw data for fixed $\dot{\delta}$ (3a) and $\delta$ (3b) and $\dot{\delta}$ (3a), respectively, and the yellow lines show their average values. The red line shows the average value of all the raw data. The black line is the same theoretical curve shown in panels a and b, normalized by its maximum value.
\label{fig:supLoops}}
\end{figure}

Finally, the light intensity is
\begin{equation}
I(t)=\frac{\dot{\varepsilon} \tau_e}{1-\lambda} \left[\frac{1}{1-\rho}
\left(e^{-t/\tau_a}- e^{-t/\tau_r}\right)-\frac{\lambda}{\lambda-\rho}
\left(e^{-t/\tau_e}-e^{-t/\tau_r}\right)\right],
\label{eq:intensity}
\end{equation}
which behaves as $I \sim  \dot{\varepsilon} t^2/2\, \tau_r$ as $t \rightarrow 0$. At large times,
with $\tau_a > \tau_e \sim\tau_r$, the dominant term in \eqref{eq:intensity} is $I\propto e^{-t/\tau_a}$, 
so $\dot{I}\sim -I/\tau_a$, a relationship seen in Figs. 3b\&c of the main text.
Figure \ref{fig:supLoops} shows plots of the solutions above. 

To find the time scales $\tau_r,\tau_e$ and $\tau_a$, we employed a least squares analysis on the 
average signal from all experiments.  The values obtained, $\tau_e \approx 0.027\,$s, 
$\tau_r \approx 0.012\,$s, and $\tau_a \approx 0.14\,$s, yield the ratios $\lambda\approx 0.19$ and 
$\rho\approx 0.09$. Thus, the prefactors within square brackets in \eqref{eq:intensity} are 
$1/(1-\rho)\approx 1.1$ and $\lambda/(\lambda-\rho)\approx 1.9$. Figure \ref{fig:supLoops}c 
compares the theoretical curve for the flash intensity with the experimental data used in Figure 3 of the main text.

\section{Finite Element Computations}

\begin{figure}[t]
\centering
\includegraphics[clip=true,width=0.9\columnwidth]{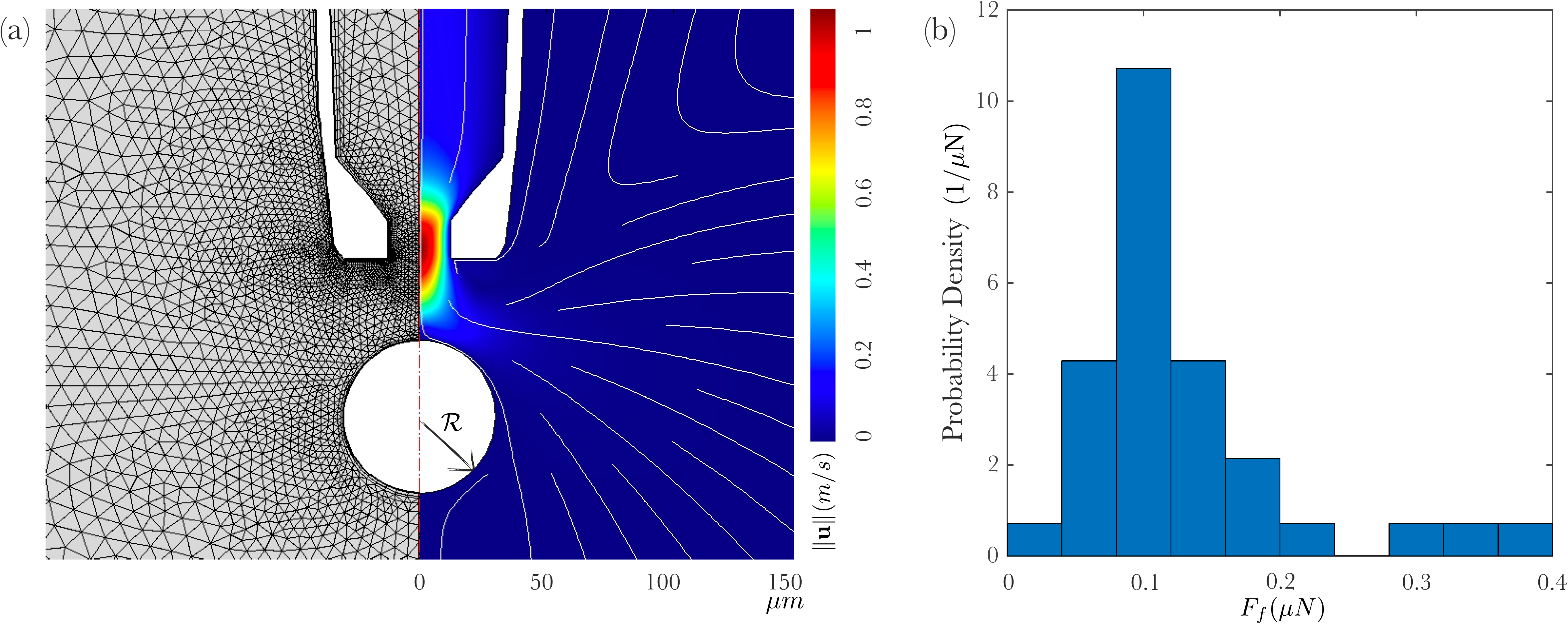}
\caption{(a) Numerical simulation of the fluid flow around an organism, modeled as a sphere: (left) mesh, (right) velocity magnitude. 
(b) Histogram of the threshold force for light production.
\label{fig:comsol}}
\end{figure}

We performed experiments and counterpart numerical computations for $35$ cells to estimate the force 
required for light production. The steady-state axisymmetric Navier-Stokes equations were solved 
numerically with the finite element software COMSOL \cite{com} to obtain the flow from a pipette impinging
on a cell. Figure \ref{fig:comsol}a shows a close up of the 
geometry employed. The geometry of the dinoflagellate is simplified to a sphere of radius $\mathcal{R}$, positioned 
a distance $H$ from the outlet of the pipette. The computational domain was chosen to be sufficiently large that
the presence of the domain boundaries did not affect the calculations. The inner diameter of the micropipette 
nozzle was set at $25\,\mu \,$m, with a flow rate $Q= 1\,$ml/h, resulting in a fluid exit speed from the
micropipette of $V\sim 1\,$m/s. Based on the actual size of the organisms, and its distance from the pipette, 
we performed the simulations for an average value of $\mathcal{R}=30\, \mu$m. The computations were found to be
insensitive to changes in $\mathcal{R}$ within our experimental values. The values of $H$ varied 
between $9$ and $75\,\mu$m. 

We compared PIV measurements of the flow created by the submerged jet in the absence of the dinoflagellate to the 
flow field computed within COMSOL, and found a good agreement. After this validation, we computed the fluid flow 
in the presence of the sphere (see Fig. \ref{fig:comsol}) and determined the mechanical forces exerted on the 
surface of the dinoflagellate (sphere) by integrating the stress over its surface. By synthesizing these 
results with the experimental thresholds for light production we obtain in Figure \ref{fig:comsol}b 
the probability distribution of the threshold force for bioluminescent flashes. The distribution peaks at 
$F_f \sim 0.1 \,\mu$N. This value is consistent with the estimation based on the dynamic pressure $\Sigma_f$ 
presented in the main text.

\end{document}